\documentclass[review,12pt]{elsarticle}

\let\today\relax
\makeatletter
\def\ps@pprintTitle{%
    \let\@oddhead\@empty
    \let\@evenhead\@empty
    \def\@oddfoot{\footnotesize\itshape
         {} \hfill\today}%
    \let\@evenfoot\@oddfoot
    
    }

\usepackage{lineno,hyperref}
\modulolinenumbers[5]
\usepackage{textgreek}
\usepackage{epsfig}
\usepackage{appendix}
\usepackage{tabu}
\usepackage{geometry}
\usepackage{amsmath}
\usepackage{pdflscape}
\usepackage{rotating}
\usepackage{longtable}
\usepackage{float}
\usepackage{placeins}
\usepackage{booktabs,caption}
\usepackage{threeparttable}
\usepackage{graphicx}
\usepackage{wrapfig}
\usepackage{lscape}
\usepackage{rotating}
\usepackage{epstopdf}
\usepackage[export]{adjustbox}
\usepackage{setspace}
\journal{}

\biboptions{authoryear}

\begin{document}

\begin{frontmatter}
\title{Exploring the Interplay of Skewness and Kurtosis: Dynamics in Cryptocurrency Markets Amid the COVID-19 Pandemic}



\author[addr]{Ariston Karagiorgis\corref{cor}}
\address[addr]{Athens University of Economics and Business, Athens, Greece}

\ead{ariskaragio@aueb.gr}
\author[add2]{Antonis Ballis}
\ead{a.ballis@aston.ac.uk}
\address[add2]{Aston Business School, Aston University, Birmingham, UK}

\author[addr]{Konstantinos Drakos}
\cortext[cor]{Corresponding author}

\ead{kdrakos@aueb.gr}
\author[add4]{Christos Kallandranis}
\ead{chriskal@uniwa.gr}
\address[add4]{University of West Attica, Athens, Greece}
\begin{abstract}
We examine how skewness interacts with kurtosis within the cryptocurrency market. We show that during the \emph{COVID-19} pandemic there are more clusters of observations around the two flanks, highlighting the presence of a volatile behavior. Moreover, we document the evolvement of the interrelationship as the pandemic progresses, identifying the domination of the extremes. Our findings advance the thinking that by exploiting the interrelationship between the two higher moments of cryptocurrencies, investors and researchers can have in their arsenal an additional analytic tool.

\end{abstract}

\begin{keyword}
%\texttt{elsarticle.cls}\sep \LaTeX\sep 
Bitcoin\sep Cryptocurrencies\sep COVID-19\sep Higher Moments \sep Skewness \sep Kurtosis. 
%\MSC[2020] 00-01\sep  99-00
\end{keyword}

\end{frontmatter}
\clearpage

\section{Introduction}

\par Undoubtedly, the coronavirus pandemic (COVID-19) as a textbook case of an exogenous shock, imposed a severe advere severely effect on financial markets. The complexity of today's financial instruments and assets, like cryptocurrencies, is by no means comparable to the ones existing during the  pandemic of 1919. The launch of Bitcoin (\cite{nakamoto_bitcoin_nodate})  was followed by an explosive expansion of the cryptocurrency market over the course of the past decade. What this essentially means is that this expansion consisted not only of an immense increase in the number of traded cryptocurrencies, but also by a rather significant inflow of funds in the newly born cryptocurrency market. 

\par  Therefore,  this erratic market/type of assets have managed to capture the attention of academia, resulting into a growing literature. Examples include \cite{urquhart_is_2019} where the hedging potential of the market was examined in their study, whereas \cite{urquhart_inefficiency_2016} and \cite{wei_liquidity_2018} had market efficiency under scrutiny. Price dynamics of the industry were investigated by \cite{phillip_new_2018}, while the area of interest of the research of \cite{cheah_speculative_2015} and \cite{corbet_datestamping_2018}  is asset pricing bubbles. Moreover, other topics of incterest include the herding behavior exhibited at the market (\cite{bouri_herding_2019}; \cite{ballis_testing_2020}) and its high volatility (\cite{feng_informed_2018}; \cite{katsiampa_volatility_2017}).

\par On the other hand, higher moments of this particular market has not gathered the level of analysis expected. As it is broadly discussed in the traditional financial literature,  returns of several financial asset classes exhibit deviations from Gaussianity (i.e. \cite{campbell1997econometrics}). \emph{Skewness} captures the distortion, in comparison to a normal distribution, separately for each tail, demonstrating the possibility for extreme events occurence and whether it's a positive or negative one. \emph{Kurtosis} on the other hand measures the extreme values in either tail, describing the shape of a distribution's tails in contrast to its overall shape.  What is less known however, especially within finance, is that \emph{Kurtosis} and \emph{Skewness} are correlated. \cite{pearson_ix_1916} and \cite{klaassen_squared_2000} proved that their interrelationship obeys well-defined rules, yet not exact.

\par Therefore, this research aims at contributing and expanding the literature, by investigating how the \emph{Skewness}-\emph{Kurtosis} interrelationship is affected by such an extreme exogenous event, as COVID-19 following the methodology of \cite{karagiorgisa2022skewness}. The two higher moments are key risk factors and often neglected parameters across the financial returns literature, let alone cryptocurrencies. In particular, we utilise the top 50 cryptocurrencies, according to market capitalization, over an 18-month period (nine months before and after the outbreak of COVID-19, as formally defined by the World Health Organization). These top 50 cryptocurrencies cover over the 95\% of the overall market.

\par The rest of this paper is organized as follows. Section 2 presents a prompt literature review. Sections 3 describes the dataset. Section 4 covers the empirical methodology. Section 5 reports the empirical findings, while Section 6 presents the conclusions of this study.

\section{Literature review}

\par It is noted within the literature (\cite{wilkins_note_1944}; \cite{groeneveld_measuring_1984}; \cite{macgillivray_relationships_1988}) that in systems characterized by disorder and deviation from normality, the \emph{S-K} plane is in general compatible with a parabolic form, but the precise structure is contingent on various factors and almost certainly includes case-specific components (\cite{alberghi2002relationship}). \cite{vargo_moment-ratio_2017} depicted the \emph{Skewness}-\emph{Kurtosis} relationship, for the vast majority of the known distributions, offering the methodology for the appropriate selection of a distribution based on empirical data. This interrelationship has been investigated in various physical phenomena (\cite{schopflocher2005relationship}, \cite{sattin_about_2009}, \cite{cristelli_universal_2012}) and in a very restricted set of asset classes. In such unstable systems this interrelationship gives rise to a \emph{S-K} plane that conveys rich information about the joint realization of \emph{Kurtosis} and \emph{Skewness} and the subsequent obedience or deviation from normality of the underlying returns' generation process. \cite{karagiorgisa2022skewness} investigated the interrelationship of the two higher moments for hedge funds returns, concluding that there is a structural relationship offering valuable insight about the differentiation of behavior between the various investment strategies. Moreover, \cite{karagiorgis2023skewness} extended the analysis within the crypto universe and researched whether the type and/or infrastracture behind each cryptoccurency reshapes the relationship.

investigate whether the type and
the infrastructure of the cryptocurrency, as well as the period under examination,
alter the architecture of the plane, finding that the squared Skewness of tokens substantially
lowers the slope of Kurtosis, while the same applies to the earlier era
8
of the market.

\par \cite{jia_higher_2021} analyzing the cross-sectional return predictability of the higher moments of 84 cryptocurrencies showcase a positive relationship between kurtosis and volatility related to future returns, while the predictability of returns for skewness is found to be negative. Building upon the aforementioned higher moments literature, we are exploiting parts of the methodology, employing it in the cryptocurrency market. Undoubtedly, cryptocurrencies due to their structure and nature, are exposed to excess skewness and kurtosis and the risks deriving from them.  

\par In his study, \cite{pearson_ix_1916} established equation \ref{eq:pearson}  as the lower bound of the \emph{S-K} plane, while \cite{schopflocher2005relationship} and later \cite{sattin_about_2009} concluded to a general form of a quadratic format with equation \ref{eq:quadratic} :

\begin{equation}\label{eq:pearson} 
K\geq S^2 +1
\end{equation}

\begin{equation}\label{eq:quadratic} 
K= A\times S^2 +B
\end{equation}

\par In their analysis, \cite{klaassen_squared_2000} approximated the lower bound of the \emph{Skewness}-\emph{Kurtosis} relationship with the formulation of equation \ref{eq:klaasen}: 

\begin{equation}\label{eq:klaasen} 
K\geq S^2 +\frac{186}{125}
\end{equation}

\par In a more recent study  \cite{cristelli_universal_2012} attempted to compare financial markets with physical phenomena (i.e., earthquakes), taking as an example the S\&P 500. The core incentive was to identify if their respective higher moments co-behave similarly and whether a universal power law (equation \ref{eq:cristelli}) can be established: 

\begin{equation}\label{eq:cristelli} 
K= N^{1/3}\times S^{4/3}
\end{equation}

%\par Furthermore, they created a validity factor $\Delta$ as described in (5) in order to test the legitimacy of the power law regime they proposed. Variable $r$ represents the return of the respective currency per week, while $r_{max}$ is the maximum return of the given period $N$. The proposed levels ($\Delta<$ 1, $\Delta>$ 10, 1$<\Delta<$ 10) for gaussian/intermediate/dominated by extreme events. 

%\begin{equation} 
%\Delta=\frac{r_{max}^4}{\sum_{i=1}^{N-1}{r^4}}
%\end{equation}

\par In the plasma physics bibliography, \cite{Kube_2016} predicted a quadratic relationship between \emph{Skewness} and \emph{Kurtosis} and fitted a regression based on equation \ref{eq:quadratic}, validating their stochastic model assumptions. A similar strategy using OLS, was followed by \cite{labit2007}. \cite{mcdonald2013skewness} exploited the ability of \emph{GB1} and  \emph{GB2} distributions to model \emph{Skewness} \& \emph{Kurtosis}, by utilizing income data. 

 %On a different work, \cite{Sattin_2009b} concluded that although their data align to a parabolic equation the existence of the quadratic relationship alone is not enough, hence only the coefficients can have explanatory value.

\section{Data}

\par Our initial dataset consists of daily data for market capitalization for each of the the top 50 cryptocurrencies, spanning the period from April 1, 2019 to September 30, 2020, with the data provided by \href{https://coinmarketcap.com/}{coinmarketcap.com}. This provides us with a panel of 23,499 day-crypto observations.  The pre COVID-19 era is defined from April 1, 2019 to December 31, 2019 and the post COVID-19 period from January 1, 2020 to September 30, 2020\footnote{We define the breaking point for the pre and post COVID-19 eras on 31 December 2019, in accordance with the timeline that WHO has provided regarding the outbreak of the COVID-19 pandemic.}.

\par \emph{Skewness} and \emph{Kurtosis} which are the main variables of our analysis, are calculated by equations \ref{eq:sk} and \ref{eq:ku} respectively, where $N$ stands for the total number of observations, $\mu$ is the sample mean, while $\sigma$ is the standard deviation and $r$ the cryptocurrency return. Furthermore,  \emph{Skewness} and \emph{Kurtosis} are calculated on weekly basis by utilising daily data.

\begin{equation}\label{eq:sk} 
S=\frac{1}{N}\sum_{i=1}^{N}\frac{(r_i-\mu)^3}{\sigma^3}
\end{equation}

\begin{equation}\label{eq:ku} 
K=\frac{1}{N}\sum_{i=1}^{N}\frac{(r_i-\mu)^4}{\sigma^4}
\end{equation}

\par Furthermore,  $\Delta$ (\cite{cristelli_universal_2012}) which is a validity factor for the power law regime is constructed, but for representation purposes is treated with $ln$ altering the corresponding thresholds ($ln\Delta<$ 0, $ln\Delta>$ 2.3, 0$<ln\Delta<$ 2.3). Finally, a dummy variable to capture the \emph{COVID-19} effect is utilised, in conjunction with its product with $Skewness^2$.

\par Table \ref{table:Descriptive} displays the univariate properties of the main variables in question of this research, and $\Delta$. \emph{Kurtosis} varies in the dataset from 1 to slightly above 5 while \emph{Skewness} takes values from -2 to 2. \emph{Skewness} has its normality area between the 25th and 50th percentile, while \emph{Kurtosis} around the 75th percentile.

\begin{center}
***** Table \ref{table:Descriptive}*****
\end{center}

\section{Methodology}

\par Building up on the literature, we are exploiting parts of the methodology, while expanding and applying it in the cryptocurrency market during an external event such as \emph{COVID-19}. Indisputably, cryptocurrencies due to their structure and nature, are exposed to excess \emph{Skewness} and \emph{Kurtosis} and the risks derived from them. 

\par Enhancing equations \ref{eq:pearson}\&\ref{eq:quadratic} that describe the \emph{S-K plane}, we develop estimation models relying on random effects and ordinary least squares (OLS). \emph{Kurtosis} is the dependent variable, with $Skewness^2 (S^2)$ the independent:
\begin{equation}
K_{i,t}=\beta_0 S_{i,t}^2+\epsilon_{i,t}
\end{equation}
%\begin{equation} H_0: \beta_0=0 \end{equation}
\par  Following, we include \emph{Skewness} ($S$): 
\begin{equation}
K_{i,t}= \beta_0 S_{i,t} + \beta_1 S_{i,t}^2+\epsilon_{i,t}
\end{equation}
%\begin{equation}H_0: \beta_0,\beta_1=0 \end{equation}

\par The next model incorporates an interaction term, defined as the the product of COVID-19 dummy ($D$) with $Skewness^2$: 
\begin{equation}\label{eq:9th}
K_{i,t}= \beta_0 S_{i,t}+\beta_1 S_{i,t}^2 +{\beta_2 S_{i,t}^2} D_{i,t} +\epsilon_{i,t}
\end{equation}

\par The hypothesis tested is for zero interaction effect:
\begin{equation}H_0: \beta_2=0 \end{equation}

\par The final estimation additionally includes the COVID-19 dummy:

\begin{equation}
K_{i,t}= \beta_0 S_{i,t}+\beta_1 S_{i,t}^2 +{\beta_2 S_{i,t}^2} D_{i,t}+ \beta_3 D_{i,t} +\epsilon_{i,t}
\end{equation}
%\begin{equation}H_0: \beta_0,\beta_1,\beta_2=0 \end{equation}

\section{Empirical Results}

 \par
\begin{landscape}

\begin{figure}[!]
\section*{Figures}
   \begin{center}
        \caption{Fitted Skewness Kurtosis interrelationship}
         \label{fig:plot1}
        \includegraphics[width=1.3\textwidth]{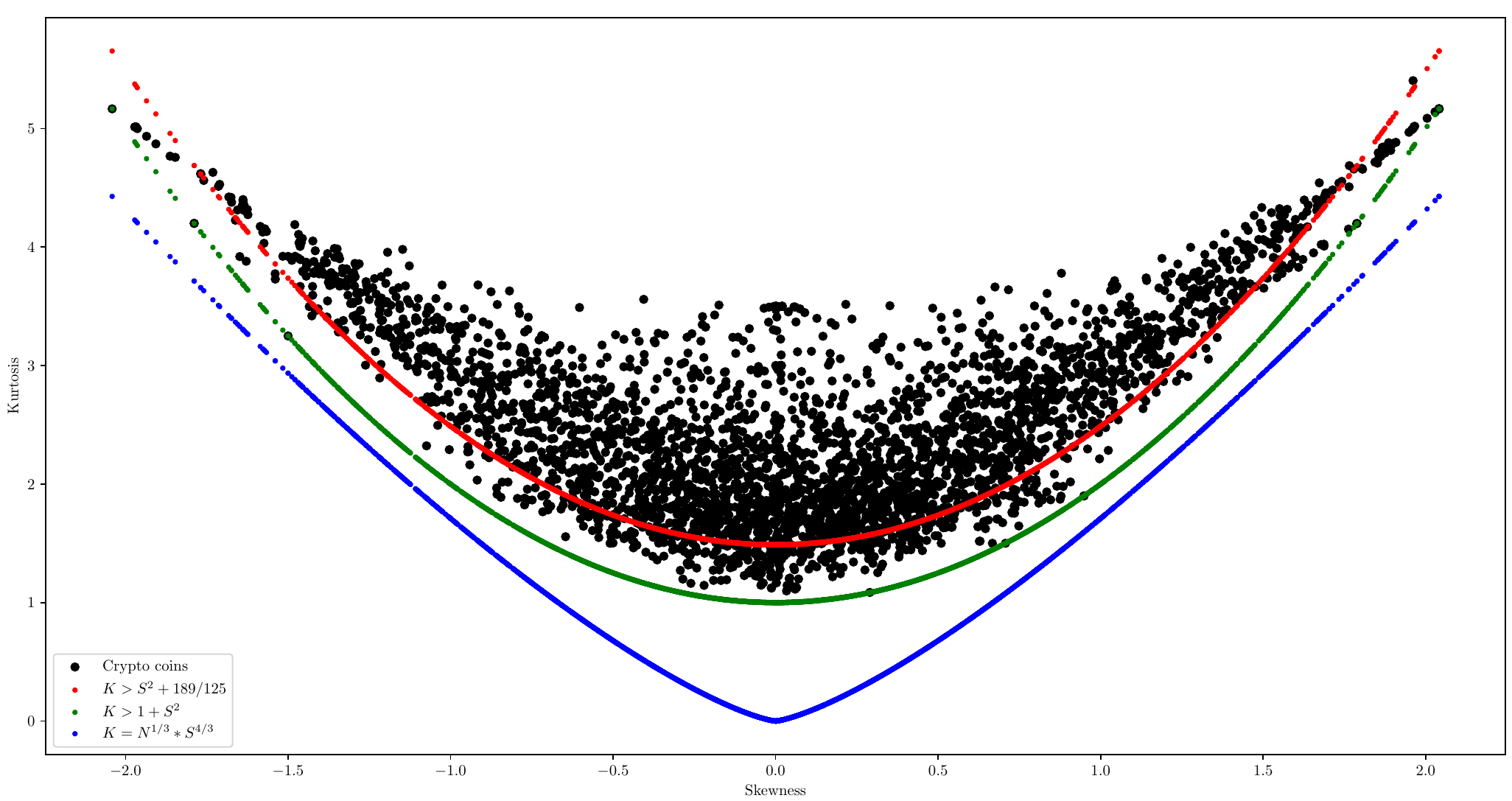}
    \end{center}    
    \end{figure}
\end{landscape}

 In Figure \ref{fig:plot1} we demonstrate on the cryptocurrencies, the three equations proposed from the bibliography. It is clear that the majority of the observations lie outside the K=3 \& S=0 area, displaying non normal distribution characteristics as anticipated. Moreover, as expected the cryptocurrency returns exhibit \emph{Kurtosis} well below the normality thresholds, demonstrating a severe fat tail risk. Regarding \emph{Skewness}, it appears that it is equally possible for the returns to be within either tail, positive or negative. The initial lower bound inequality established by \cite{pearson_ix_1916} indeed seems to be the extremum of the scatter points. Moreover, the equation suggested by \cite{klaassen_squared_2000} fits the data profoundly, in contrast with the proposed power law of \cite{cristelli_universal_2012} which assumes a universal relation amid the higher moments.

\begin{center}
***** Figure \ref{fig:plot1}*****
\end{center}

\par
\begin{landscape}
\begin{figure}[!]
   \begin{center}
        \caption{Pre vs Post COVID-19 comparison}
         \label{fig:era}
        \includegraphics[width=1.3\textwidth]{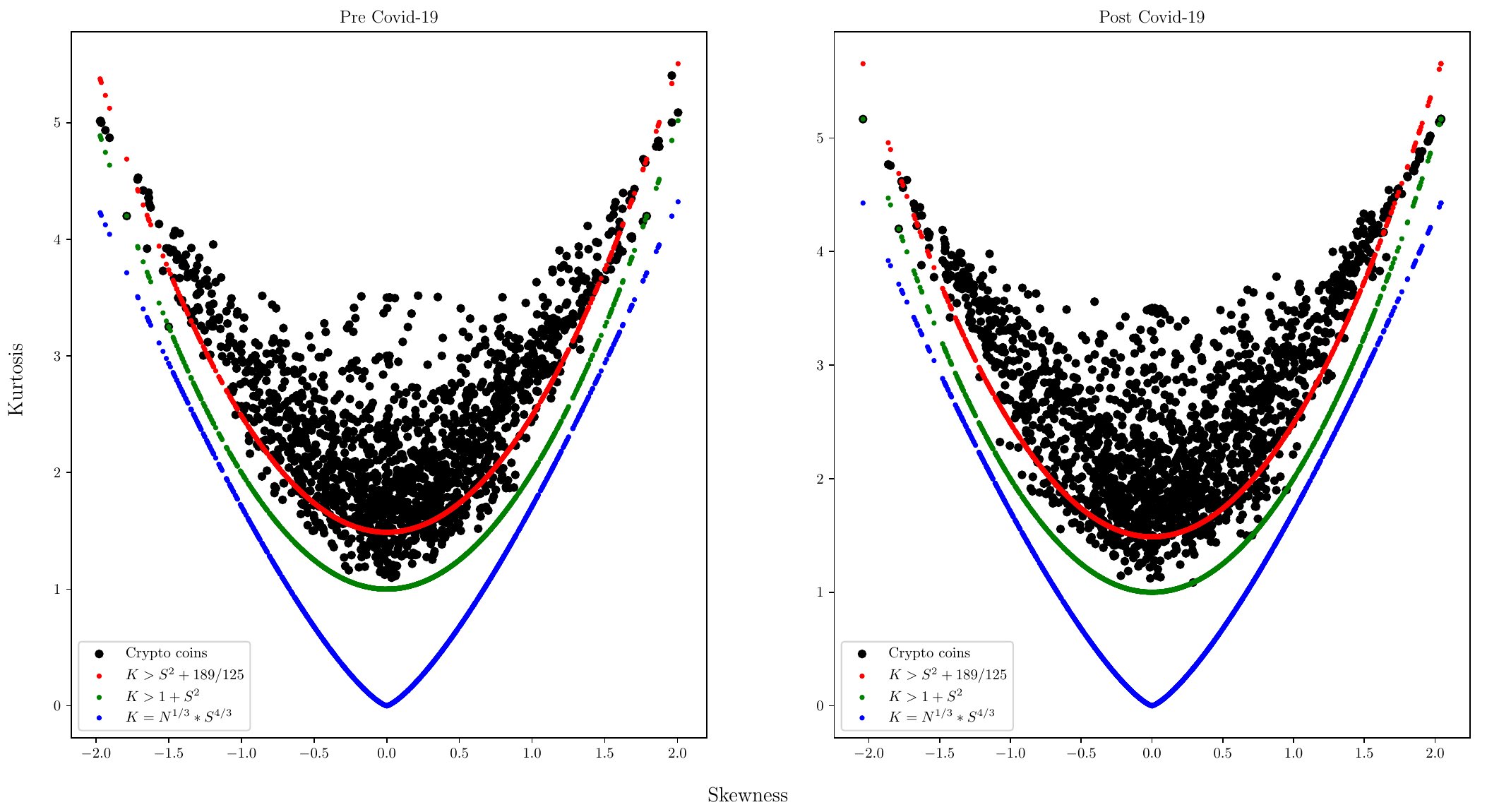}
    \end{center}    
    \end{figure}
\end{landscape}

 Furthermore, Figure \ref{fig:era} depicts the \emph{S-K} plane on the span of 9 months prior the announcement on the left panel and up to 9 months after on the right. Although the two scatter plots do resemble one another, there are some differences between them. On the post COVID-19 announcement era, there are substantially more clusters of observations around the two flanks, with either positive or negative \emph{Skewness} and \emph{Kurtosis} above 4, highlighting the volatile behavior.

\begin{center}
***** Figure \ref{fig:era}*****
\end{center}
\par
\begin{landscape}
\begin{figure}[!]
   \begin{center}
        \caption{\emph{Skewness-Kurtosis}- $\Delta$ - Time }
         \label{fig:4d}
        \includegraphics[width=1.3\textwidth]{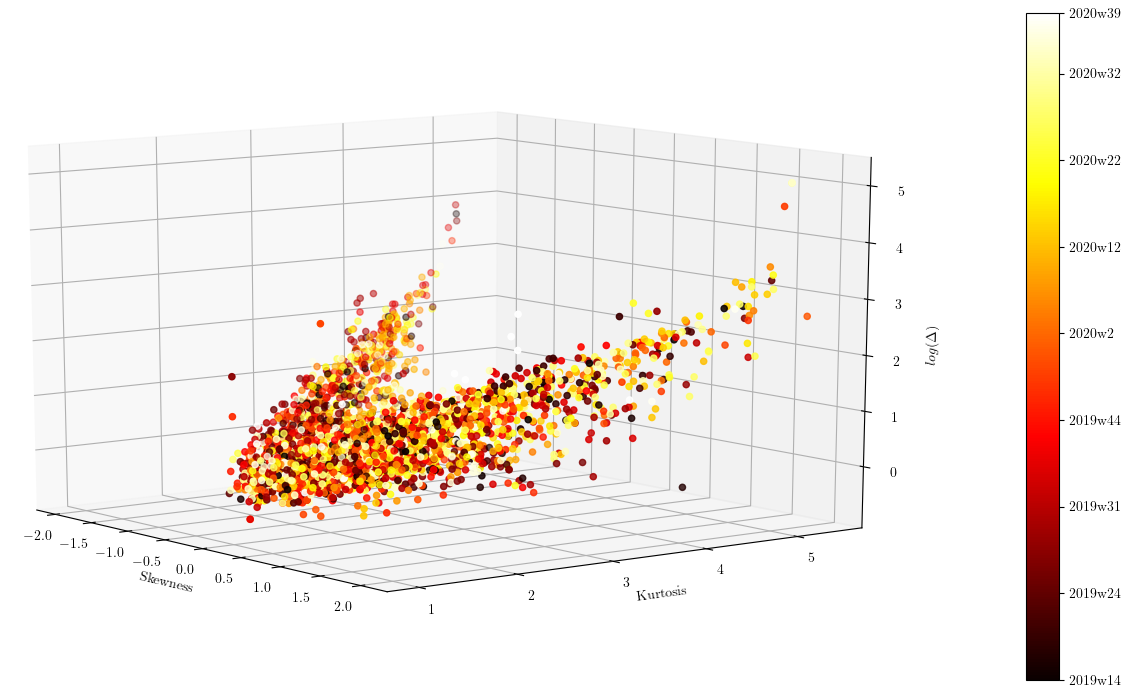}
    \end{center}    
    \end{figure}
\end{landscape}

 Figure \ref{fig:4d} exhibits the relation between skewness-kurtosis-time and factor $\Delta$. The initial threshold of $\Delta$ for normality is 0, the intermediate state is up to 2.3 and above that is the area where extremes drive the distribution. As showcased earlier in Table 1, more than 25\% of the returns can be observed on the latter area and less than 25\% within the normality region. While observations of high $\Delta$  can be observed across the \emph{S-K} plane, as it is anticipated the most extreme values can be found mainly on the two tails. Moreover, it appears that  $\Delta$ gradually increases with \emph{Kurtosis} and there is a cluster of values mainly above the area where kurtosis reaches 4. \emph{Skewness} can be either positive or negative, but well away from normality. 

\par The heat map allows us to pin down the COVID-19 effect and on the behavior of cryptocurrencies returns. The darker points which represent 2019 values appear to be around or near areas where normality is defined; \emph{Skewness} near zero and \emph{Kurtosis} around 3 and with lower $\Delta$ values. It is evident that as time progresses and the scatter points get lighter, which are well in 2020 and the pandemic has evolved along with uncertainty, $\Delta$ tends to assume higher values. The weeks of September 2020 which are represented by almost white color, are heavily positively skewed but around the normal area of kurtosis and vast $\Delta$ values. This behavior could be attributed to positive news i.e. about a vaccine.
\begin{center}
***** Figure \ref{fig:4d}*****
\end{center}

\par  We proceed to formally examine the relationship that already has been established. Table \ref{table:quadratic} displays four different estimations of the quadratic equation, using random effects. On all occasions, $Skewness^2$ have the same amplifying effect on \emph{Kurtosis}, while it is significant at all conventional levels. Coefficient and significance level is in accordance to \cite{karagiorgis2023skewness} and their research on the entirety of the cryptocurrencies' universe in a bigger time frame.  
On the other hand, \emph{Skewness} appears to have a minor negative effect on \emph{Kurtosis} and statistical significance at 5\% in contrast to the aforementioned investigation.

\par On equation \ref{eq:9th}, the interaction of $Skewness^2$ with the \emph{COVID-19} dummy, provides an additional effect on \emph{Kurtosis}, attributed to the deteriorating conditions in the markets due to the pandemic. The $H_0$ for zero interaction effect is rejected.  Moreover, while in the final estimation the slope of the regression line remains unaffected by the pandemic, this is not the case for the mean value of \emph{Kurtosis}. \emph{COVID-19} substantially increases it, with significance at 1\%, due to the extremes produced during this period. We also test for total statistical insignificance of the model, with the hypothesis being rejected.

\begin{center}
***** Table \ref{table:quadratic}*****
\end{center}

\par As an additional check we estimate the same four models with OLS as shown in Table \ref{table:ols}. All the variables demonstrate the same behavior as before, while the \emph{COVID-19} interaction coefficient has slightly higher coefficient in (9). The F tests provide sufficient evidence for the significance of the model.

\begin{center}
***** Table \ref{table:ols}*****
\end{center}

\section{Conclusions}

\par The aim of this study was to explore how skewness interacts with kurtosis within the cryptocurrency market. Both the higher moments portrait the tail risks. Our analysis utilised daily data from the top 50 cryptocurrencies that represent, on average, over 95\% of the overall market, covering the period from April 2019 to September 2020. Our results indicate that during the COVID-19 pandemic era there are more clusters of observations around the two flanks, highlighting the volatile behavior. Furthermore, we document the evolvement of the interrelationship as the pandemic progresses, identifying the domination of the extremes. Finally, we have validated the significance of the quadratic equation via three different models. Overall, our findings suggest that by exploiting the interrelationship between the two higher moments of cryptocurrencies, investors and researchers can have an additional analytic tool.

\cleardoublepage
\section*{References}

\bibliography{mybibfile}

\begin{thebibliography}{27}
\expandafter\ifx\csname natexlab\endcsname\relax\def\natexlab#1{#1}\fi
\providecommand{\url}[1]{\texttt{#1}}
\providecommand{\href}[2]{#2}
\providecommand{\path}[1]{#1}
\providecommand{\DOIprefix}{doi:}
\providecommand{\ArXivprefix}{arXiv:}
\providecommand{\URLprefix}{URL: }
\providecommand{\Pubmedprefix}{pmid:}
\providecommand{\doi}[1]{\href{http://dx.doi.org/#1}{\path{#1}}}
\providecommand{\Pubmed}[1]{\href{pmid:#1}{\path{#1}}}
\providecommand{\bibinfo}[2]{#2}
\ifx\xfnm\relax \def\xfnm[#1]{\unskip,\space#1}\fi
%Type = Article
\bibitem[{Alberghi et~al.(2002)Alberghi, Maurizi and
  Tampieri}]{alberghi2002relationship}
\bibinfo{author}{Alberghi, S.}, \bibinfo{author}{Maurizi, A.},
  \bibinfo{author}{Tampieri, F.}, \bibinfo{year}{2002}.
\newblock \bibinfo{title}{Relationship between the vertical velocity skewness
  and kurtosis observed during sea-breeze convection}.
\newblock \bibinfo{journal}{Journal of Applied Meteorology}
  \bibinfo{volume}{41}, \bibinfo{pages}{885--889}.
%Type = Article
\bibitem[{Ballis and Drakos(2020)}]{ballis_testing_2020}
\bibinfo{author}{Ballis, A.}, \bibinfo{author}{Drakos, K.},
  \bibinfo{year}{2020}.
\newblock \bibinfo{title}{Testing for herding in the cryptocurrency market}.
\newblock \bibinfo{journal}{Finance Research Letters} \bibinfo{volume}{33},
  \bibinfo{pages}{101210}.
\newblock \DOIprefix\doi{10.1016/j.frl.2019.06.008}.
%Type = Article
\bibitem[{Bouri et~al.(2019)Bouri, Gupta and Roubaud}]{bouri_herding_2019}
\bibinfo{author}{Bouri, E.}, \bibinfo{author}{Gupta, R.},
  \bibinfo{author}{Roubaud, D.}, \bibinfo{year}{2019}.
\newblock \bibinfo{title}{Herding behaviour in cryptocurrencies}.
\newblock \bibinfo{journal}{Finance Research Letters} \bibinfo{volume}{29},
  \bibinfo{pages}{216--221}.
\newblock \URLprefix
  \url{https://www.sciencedirect.com/science/article/pii/S1544612318303647},
  \DOIprefix\doi{10.1016/j.frl.2018.07.008}.
%Type = Book
\bibitem[{Campbell et~al.(1997)Campbell, Champbell, Campbell, Lo, Lo and
  MacKinlay}]{campbell1997econometrics}
\bibinfo{author}{Campbell, J.Y.}, \bibinfo{author}{Champbell, J.J.},
  \bibinfo{author}{Campbell, J.W.}, \bibinfo{author}{Lo, A.W.},
  \bibinfo{author}{Lo, A.W.}, \bibinfo{author}{MacKinlay, A.C.},
  \bibinfo{year}{1997}.
\newblock \bibinfo{title}{The econometrics of financial markets}.
\newblock \bibinfo{publisher}{princeton University press}.
%Type = Article
\bibitem[{Cheah and Fry(2015)}]{cheah_speculative_2015}
\bibinfo{author}{Cheah, E.T.}, \bibinfo{author}{Fry, J.}, \bibinfo{year}{2015}.
\newblock \bibinfo{title}{Speculative bubbles in {Bitcoin} markets? {An}
  empirical investigation into the fundamental value of {Bitcoin}}.
\newblock \bibinfo{journal}{Economics Letters} \bibinfo{volume}{130},
  \bibinfo{pages}{32--36}.
\newblock \DOIprefix\doi{10.1016/j.econlet.2015.02.029}.
%Type = Article
\bibitem[{Corbet et~al.(2018)Corbet, Lucey and
  Yarovaya}]{corbet_datestamping_2018}
\bibinfo{author}{Corbet, S.}, \bibinfo{author}{Lucey, B.},
  \bibinfo{author}{Yarovaya, L.}, \bibinfo{year}{2018}.
\newblock \bibinfo{title}{Datestamping the {Bitcoin} and {Ethereum} bubbles}.
\newblock \bibinfo{journal}{Finance Research Letters} \bibinfo{volume}{26},
  \bibinfo{pages}{81--88}.
\newblock \DOIprefix\doi{10.1016/j.frl.2017.12.006}.
%Type = Article
\bibitem[{Cristelli et~al.(2012)Cristelli, Zaccaria and
  Pietronero}]{cristelli_universal_2012}
\bibinfo{author}{Cristelli, M.}, \bibinfo{author}{Zaccaria, A.},
  \bibinfo{author}{Pietronero, L.}, \bibinfo{year}{2012}.
\newblock \bibinfo{title}{Universal relation between skewness and kurtosis in
  complex dynamics}.
\newblock \bibinfo{journal}{Physical Review E} \bibinfo{volume}{85},
  \bibinfo{pages}{066108}.
\newblock \DOIprefix\doi{10.1103/PhysRevE.85.066108}.
%Type = Article
\bibitem[{Feng et~al.(2018)Feng, Wang and Zhang}]{feng_informed_2018}
\bibinfo{author}{Feng, W.}, \bibinfo{author}{Wang, Y.}, \bibinfo{author}{Zhang,
  Z.}, \bibinfo{year}{2018}.
\newblock \bibinfo{title}{Informed trading in the {Bitcoin} market}.
\newblock \bibinfo{journal}{Finance Research Letters} \bibinfo{volume}{26},
  \bibinfo{pages}{63--70}.
\newblock \DOIprefix\doi{10.1016/j.frl.2017.11.009}.
%Type = Article
\bibitem[{Groeneveld and Meeden(1984)}]{groeneveld_measuring_1984}
\bibinfo{author}{Groeneveld, R.A.}, \bibinfo{author}{Meeden, G.},
  \bibinfo{year}{1984}.
\newblock \bibinfo{title}{Measuring {Skewness} and {Kurtosis}}.
\newblock \bibinfo{journal}{Journal of the Royal Statistical Society: Series D
  (The Statistician)} \bibinfo{volume}{33}, \bibinfo{pages}{391--399}.
\newblock \URLprefix
  \url{https://rss.onlinelibrary.wiley.com/doi/abs/10.2307/2987742},
  \DOIprefix\doi{10.2307/2987742}. \bibinfo{note}{\_eprint:
  https://rss.onlinelibrary.wiley.com/doi/pdf/10.2307/2987742}.
%Type = Article
\bibitem[{Jia et~al.(2021)Jia, Liu and Yan}]{jia_higher_2021}
\bibinfo{author}{Jia, Y.}, \bibinfo{author}{Liu, Y.}, \bibinfo{author}{Yan,
  S.}, \bibinfo{year}{2021}.
\newblock \bibinfo{title}{Higher moments, extreme returns, and cross–section
  of cryptocurrency returns}.
\newblock \bibinfo{journal}{Finance Research Letters} \bibinfo{volume}{39},
  \bibinfo{pages}{101536}.
\newblock \URLprefix
  \url{https://www.sciencedirect.com/science/article/pii/S1544612320303135},
  \DOIprefix\doi{10.1016/j.frl.2020.101536}.
%Type = Article
\bibitem[{Karagiorgis and Drakos(2022)}]{karagiorgisa2022skewness}
\bibinfo{author}{Karagiorgis, A.}, \bibinfo{author}{Drakos, K.},
  \bibinfo{year}{2022}.
\newblock \bibinfo{title}{The skewness-kurtosis plane for non-gaussian systems:
  The case of hedge fund returns}.
\newblock \bibinfo{journal}{Journal of International Financial Markets,
  Institutions and Money} , \bibinfo{pages}{101639}.
%Type = Article
\bibitem[{Katsiampa(2017)}]{katsiampa_volatility_2017}
\bibinfo{author}{Katsiampa, P.}, \bibinfo{year}{2017}.
\newblock \bibinfo{title}{Volatility estimation for {Bitcoin}: {A} comparison
  of {GARCH} models}.
\newblock \bibinfo{journal}{Economics Letters} \bibinfo{volume}{158},
  \bibinfo{pages}{3--6}.
\newblock \DOIprefix\doi{10.1016/j.econlet.2017.06.023}.
%Type = Article
\bibitem[{Klaassen et~al.(2000)Klaassen, Mokveld and van
  Es}]{klaassen_squared_2000}
\bibinfo{author}{Klaassen, C.A.J.}, \bibinfo{author}{Mokveld, P.J.},
  \bibinfo{author}{van Es, B.}, \bibinfo{year}{2000}.
\newblock \bibinfo{title}{Squared skewness minus kurtosis bounded by 186/125
  for unimodal distributions}.
\newblock \bibinfo{journal}{Statistics \& Probability Letters}
  \bibinfo{volume}{50}, \bibinfo{pages}{131--135}.
\newblock \DOIprefix\doi{10.1016/S0167-7152(00)00090-0}.
%Type = Article
\bibitem[{Kube et~al.(2016)Kube, Theodorsen, Garcia, LaBombard and
  Terry}]{Kube_2016}
\bibinfo{author}{Kube, R.}, \bibinfo{author}{Theodorsen, A.},
  \bibinfo{author}{Garcia, O.E.}, \bibinfo{author}{LaBombard, B.},
  \bibinfo{author}{Terry, J.L.}, \bibinfo{year}{2016}.
\newblock \bibinfo{title}{Fluctuation statistics in the scrape-off layer of
  alcator c-mod}.
\newblock \bibinfo{journal}{Plasma Physics and Controlled Fusion}
  \bibinfo{volume}{58}, \bibinfo{pages}{054001}.
\newblock \DOIprefix\doi{10.1088/0741-3335/58/5/054001}.
%Type = Article
\bibitem[{Labit et~al.(2007)Labit, Furno, Fasoli, Diallo, M\"uller, Plyushchev,
  Podest\`a and Poli}]{labit2007}
\bibinfo{author}{Labit, B.}, \bibinfo{author}{Furno, I.},
  \bibinfo{author}{Fasoli, A.}, \bibinfo{author}{Diallo, A.},
  \bibinfo{author}{M\"uller, S.H.}, \bibinfo{author}{Plyushchev, G.},
  \bibinfo{author}{Podest\`a, M.}, \bibinfo{author}{Poli, F.M.},
  \bibinfo{year}{2007}.
\newblock \bibinfo{title}{Universal statistical properties of drift-interchange
  turbulence in torpex plasmas}.
\newblock \bibinfo{journal}{Phys. Rev. Lett.} \bibinfo{volume}{98},
  \bibinfo{pages}{255002}.
\newblock \DOIprefix\doi{10.1103/PhysRevLett.98.255002}.
%Type = Article
\bibitem[{MacGillivray and Balanda(1988)}]{macgillivray_relationships_1988}
\bibinfo{author}{MacGillivray, H.}, \bibinfo{author}{Balanda, K.},
  \bibinfo{year}{1988}.
\newblock \bibinfo{title}{The relationships between skewness and kurtosis}.
\newblock \bibinfo{journal}{Australian Journal of Statistics}
  \bibinfo{volume}{30}, \bibinfo{pages}{319--337}.
\newblock \URLprefix
  \url{https://onlinelibrary.wiley.com/doi/abs/10.1111/j.1467-842X.1988.tb00626.x?casa_token=QVbvI6oCgMgAAAAA%3AKXGYsND5fZyzGgI-gbTteTA6_Pr0ucwlLMTN6bo2jYyM9wcdN2RAep76c1hgJqqwMjm3nO6mlO7v-DY},
  \DOIprefix\doi{https://doi.org/10.1111/j.1467-842X.1988.tb00626.x}.
%Type = Article
\bibitem[{McDonald et~al.(2013)McDonald, Sorensen and
  Turley}]{mcdonald2013skewness}
\bibinfo{author}{McDonald, J.B.}, \bibinfo{author}{Sorensen, J.},
  \bibinfo{author}{Turley, P.A.}, \bibinfo{year}{2013}.
\newblock \bibinfo{title}{Skewness and kurtosis properties of income
  distribution models}.
\newblock \bibinfo{journal}{Review of Income and Wealth} \bibinfo{volume}{59},
  \bibinfo{pages}{360--374}.
\newblock \URLprefix
  \url{https://onlinelibrary.wiley.com/doi/abs/10.1111/j.1475-4991.2011.00478.x},
  \DOIprefix\doi{https://doi.org/10.1111/j.1475-4991.2011.00478.x}.
%Type = Techreport
\bibitem[{Nakamoto(2008)}]{nakamoto_bitcoin_nodate}
\bibinfo{author}{Nakamoto, S.}, \bibinfo{year}{2008}.
\newblock \bibinfo{title}{Bitcoin: {A} {Peer}-to-{Peer} {Electronic} {Cash}
  {System}}.
\newblock \bibinfo{type}{Technical Report}. Manubot.
\newblock \URLprefix \url{https://git.dhimmel.com/bitcoin-whitepaper/}.
  \bibinfo{note}{publication Title: Manubot}.
%Type = Article
\bibitem[{Pearson(1916)}]{pearson_ix_1916}
\bibinfo{author}{Pearson, K.}, \bibinfo{year}{1916}.
\newblock \bibinfo{title}{{IX}. {Mathematical} contributions to the theory of
  evolution.—{XIX}. {Second} supplement to a memoir on skew variation}.
\newblock \bibinfo{journal}{Philosophical Transactions of the Royal Society of
  London. Series A, Containing Papers of a Mathematical or Physical Character}
  \bibinfo{volume}{216}, \bibinfo{pages}{429--457}.
\newblock \DOIprefix\doi{10.1098/rsta.1916.0009}.
%Type = Article
\bibitem[{Phillip et~al.(2018)Phillip, Chan and Peiris}]{phillip_new_2018}
\bibinfo{author}{Phillip, A.}, \bibinfo{author}{Chan, J.S.K.},
  \bibinfo{author}{Peiris, S.}, \bibinfo{year}{2018}.
\newblock \bibinfo{title}{A new look at {Cryptocurrencies}}.
\newblock \bibinfo{journal}{Economics Letters} \bibinfo{volume}{163},
  \bibinfo{pages}{6--9}.
\newblock \URLprefix
  \url{https://www.sciencedirect.com/science/article/pii/S0165176517304731},
  \DOIprefix\doi{10.1016/j.econlet.2017.11.020}.
%Type = Article
\bibitem[{Sattin et~al.(2009)Sattin, Agostini, Cavazzana, Serianni, Scarin and
  Vianello}]{sattin_about_2009}
\bibinfo{author}{Sattin, F.}, \bibinfo{author}{Agostini, M.},
  \bibinfo{author}{Cavazzana, R.}, \bibinfo{author}{Serianni, G.},
  \bibinfo{author}{Scarin, P.}, \bibinfo{author}{Vianello, N.},
  \bibinfo{year}{2009}.
\newblock \bibinfo{title}{About the parabolic relation existing between the
  skewness and the kurtosis in time series of experimental data}.
\newblock \bibinfo{journal}{Physica Scripta} \bibinfo{volume}{79},
  \bibinfo{pages}{045006}.
\newblock \URLprefix
  \url{https://doi.org/10.1088%2F0031-8949%2F79%2F04%2F045006},
  \DOIprefix\doi{10.1088/0031-8949/79/04/045006}. \bibinfo{note}{publisher: IOP
  Publishing}.
%Type = Article
\bibitem[{Schopflocher and Sullivan(2005)}]{schopflocher2005relationship}
\bibinfo{author}{Schopflocher, T.}, \bibinfo{author}{Sullivan, P.},
  \bibinfo{year}{2005}.
\newblock \bibinfo{title}{The relationship between skewness and kurtosis of a
  diffusing scalar}.
\newblock \bibinfo{journal}{Boundary-layer meteorology} \bibinfo{volume}{115},
  \bibinfo{pages}{341--358}.
\newblock \DOIprefix\doi{https://doi.org/10.1007/s10546-004-5642-7}.
%Type = Article
\bibitem[{Urquhart(2016)}]{urquhart_inefficiency_2016}
\bibinfo{author}{Urquhart, A.}, \bibinfo{year}{2016}.
\newblock \bibinfo{title}{The inefficiency of {Bitcoin}}.
\newblock \bibinfo{journal}{Economics Letters} \bibinfo{volume}{148},
  \bibinfo{pages}{80--82}.
%Type = Article
\bibitem[{Urquhart and Zhang(2019)}]{urquhart_is_2019}
\bibinfo{author}{Urquhart, A.}, \bibinfo{author}{Zhang, H.},
  \bibinfo{year}{2019}.
\newblock \bibinfo{title}{Is {Bitcoin} a hedge or safe haven for currencies?
  {An} intraday analysis}.
\newblock \bibinfo{journal}{International Review of Financial Analysis}
  \bibinfo{volume}{63}, \bibinfo{pages}{49--57}.
\newblock \DOIprefix\doi{10.1016/j.irfa.2019.02.009}.
%Type = Incollection
\bibitem[{Vargo et~al.(2017)Vargo, Pasupathy and
  Leemis}]{vargo_moment-ratio_2017}
\bibinfo{author}{Vargo, E.}, \bibinfo{author}{Pasupathy, R.},
  \bibinfo{author}{Leemis, L.M.}, \bibinfo{year}{2017}.
\newblock \bibinfo{title}{Moment-{Ratio} {Diagrams} for {Univariate}
  {Distributions}}, in: \bibinfo{editor}{Glen, A.G.}, \bibinfo{editor}{Leemis,
  L.M.} (Eds.), \bibinfo{booktitle}{Computational {Probability}
  {Applications}}. \bibinfo{publisher}{Springer International Publishing},
  \bibinfo{address}{Cham}. International {Series} in {Operations} {Research} \&
  {Management} {Science}, pp. \bibinfo{pages}{149--164}.
\newblock \URLprefix \url{https://doi.org/10.1007/978-3-319-43317-2_12},
  \DOIprefix\doi{10.1007/978-3-319-43317-2_12}.
%Type = Article
\bibitem[{Wei(2018)}]{wei_liquidity_2018}
\bibinfo{author}{Wei, W.C.}, \bibinfo{year}{2018}.
\newblock \bibinfo{title}{Liquidity and market efficiency in cryptocurrencies}.
\newblock \bibinfo{journal}{Economics Letters} \bibinfo{volume}{168},
  \bibinfo{pages}{21--24}.
\newblock \DOIprefix\doi{10.1016/j.econlet.2018.04.003}.
%Type = Article
\bibitem[{Wilkins(1944)}]{wilkins_note_1944}
\bibinfo{author}{Wilkins, J.E.}, \bibinfo{year}{1944}.
\newblock \bibinfo{title}{A {Note} on {Skewness} and {Kurtosis}}.
\newblock \bibinfo{journal}{Annals of Mathematical Statistics}
  \bibinfo{volume}{15}, \bibinfo{pages}{333--335}.
\newblock \URLprefix \url{https://projecteuclid.org/euclid.aoms/1177731243},
  \DOIprefix\doi{10.1214/aoms/1177731243}. \bibinfo{note}{publisher: Institute
  of Mathematical Statistics}.

\end{thebibliography}
\cleardoublepage

%\section*{Tables}
%\appendix
%%%%%figures
%%%%simplescatter
%\begin{figure}[ht]
  % \begin{center}
     %   \includegraphics[width=1\textwidth]{scatter1}
        %\caption{}
         %\label{fig:fs_mi}
    %\end{center}    
   % \end{figure}
% \begin{figure}
   %\begin{center}
      %  \includegraphics[width=10.5cm,height=50cm,keepaspectratio]{Figure_11}
       % \caption{}
        % \label{fig:fs_mi}
   % \end{center}    
   % \end{figure} 

 %\begin{figure}
  % \begin{center}
    %    \includegraphics[width=14.1cm,height=50cm,keepaspectratio]{Figure_21}
       % \caption{}
        % \label{fig:fs_mi}
    %\end{center}    
   % \end{figure} \clearpage
%\setlength{\arrayrulewidth}{.3em}
\begin{landscape}\section*{Tables}
\begin{longtable}{lllllllllll}
\caption{Descriptive statistics of higher moments}
\label{table:Descriptive}\\ 
\hline

     & N     & Mean   & St.Dev & min    & max    & p1     & p25    & p50    & p75   & p99   \\\hline
%$Skewness^2$ & 3423 & 0.734  & 1.094  & 0      & 9.084  & 0      & 0.046  & 0.22   & 0.685 & 5.788 \\
Skewness   & 3423 & 0.082 & 0.734  & -2.041 & 2.041  & -1.566 & -0.389 & 0.061 & 0.581 & 1.779 \\
Kurtosis   & 3423 & 2.473  & 0.780   & 1      & 5.405 & 1.243  & 1.855  & 2.339  & 2.973 & 4.757 \\
$\Delta$   & 3409 & 116.450  & 2995.92   & 0.205      & 119178.9 & 0.376  & 1.245  & 2.868  & 10.261 & 576.588 \\
\hline
\end{longtable}
\end{landscape}

\cleardoublepage
\FloatBarrier
\scriptsize
\centering
%\captionof{table}{Random Effects}
\begin{longtable}{lcccc}
\caption{Quadratic model}
\label{table:quadratic}\\ 
\hline 
Kurtosis & (7)& (8) &(9) & (11) \\ \hline
 &  &  &&  \\
$Skewness^2$ & 0.880*** & 0.883*** & 0.855***&0.893*** \\
 & (0.0103) & (0.0104) & (0.0143)&(0.0156) \\
Skewness &  & -0.0210** & -0.0216**&-0.0204** \\
 &  & (0.0102) & (0.0102)&(0.010) \\
$Skewness^2\times$COVID-19  &  &  & 0.0464***&-0.026 \\
 &  &  & (0.0164)&(0.020) \\
 COVID-19&  &  & & 0.111***\\
 &  &  & &(0.018) \\
Constant & 1.993*** & 1.993*** & 1.994***&1.936 \\
 & (0.00925) & (0.00924) & (0.00924)&(0.023) \\
 &  &  & &  \\
 \hline
Observations & 3,423 & 3,423 & 3,423 & 3,423 \\
\begin{tabular}{@{}ll@{}}
                   R-squared & \multicolumn{1}{m{1.5cm}}{within  \ \ \ \  between   overall } \\
                 \end{tabular}  & \begin{tabular}{@{}l@{}}
                   0.684\\
                   0.610\\
                   0.672\\
                 \end{tabular} & \begin{tabular}{@{}l@{}}
                   0.684\\
                   0.611\\
                   0.672\\
                 \end{tabular} & \begin{tabular}{@{}l@{}}
                   0.685\\
                   0.610\\
                   0.673\\
                 \end{tabular} & \begin{tabular}{@{}l@{}}
                   0.688\\
                   0.596\\
                   0.676\\
                 \end{tabular} \\\hline
\multicolumn{1}{m{4cm}}{Joint test for zero interaction effects }&-&-&$X^2_{1}= 8.04 $***&$X^2_{1}= 1.71 $\\\hline
\multicolumn{1}{m{4cm}}{Joint test for zero total  effects }&$X^2_{1}= 7343.81 $***&$X^2_{2}= 7355.87 $***&$X^2_{3}= 7380.15 $***&$X^2_{4}= 7493.72 $***\\\hline
%\multicolumn{1}{m{3cm}}{Joint test for zero direct macro effects }&$X^2_{3}= 17.39 $***&-&-&-\\\hline
%Chi-squared&7280.91&3646.01&2438.37&7493.72\\
%Prob$>$F&0.0000&0.0000&0.0000&0.0000\\
 %Number of id & 50 & 50 & 50 \\ 
 \hline
\multicolumn{5}{c}{ Standard errors in parentheses} \\
\multicolumn{5}{c}{ *** p$<$0.01, ** p$<$0.05, * p$<$0.1} \\
\end{longtable}
\cleardoublepage
\centering

\begin{longtable}{lcccc} 
\caption{Quadratic model}
\label{table:ols}\\ 
\hline 
Kurtosis & (7)& (8) &(9) & (11) \\ \hline
 &  &  & & \\
$Skewness^2$ & 0.872*** & 0.876*** & 0.847****&0.886*** \\
 & (0.0104) & (0.0106) & (0.0146)& (0.0160) \\
Skewness &  & -0.0241** & -0.0248**& -0.0229** \\
 &  & (0.0106) & (0.0105)&(0.0105) \\
$Skewness^2\times$COVID-19  &  &  & 0.0495***&-0.022 \\
 &  &  & (0.0170)&(0.0210) \\
 COVID-19   &  &  & &0.109*** \\
 &  &  & &(0.01897) \\
Constant & 1.997*** & 1.997*** & 1.998***&1.940*** \\
 & (0.00953) & (0.00952) & (0.00952)& (0.0137)\\
 &  &  &  \\
 \hline
Observations & 3,423 & 3,423 & 3,423 & 3,423 \\
 R-squared & 0.672 & 0.672 & 0.673& 0.676 \\\hline
\multicolumn{1}{m{2.5cm}}{Joint test for zero interaction effects}&- & -& $F_{1,3419}=8.50$***  &  $F_{1,3418}=1.19$   \\\hline
\multicolumn{1}{m{2.5cm}}{Joint test for zero total  effects}& $F_{1,3421}=6998.33$*** & $F_{2,3420}=3506.08$*** & $F_{3,3419}=2345.35$***  &   $F_{4,3418}=1783.97$***   \\\hline 

 \hline
\multicolumn{5}{c}{ Standard errors in parentheses} \\
\multicolumn{5}{c}{ *** p$<$0.01, ** p$<$0.05, * p$<$0.1} \\
\end{longtable}

\FloatBarrier

\end{document}